\documentclass[twocolumn,aps,pra,superscriptaddress, floatfix]{revtex4-1}

\usepackage[utf8]{inputenc}
\usepackage{graphicx}
\usepackage{natbib}
\usepackage{url}
\usepackage{textcase}
\usepackage{bm}
\usepackage{bbm}
\usepackage{times}
\usepackage{amsmath, amsthm}
\usepackage{physics}
\usepackage{hyperref}
\usepackage{dcolumn}
\usepackage[english]{babel}
\usepackage[utf8]{inputenc}
\usepackage{verbatim}
\usepackage{xcolor}
\usepackage{color,soul}
\usepackage{siunitx}
\usepackage{lineno}
\usepackage{tabularray}
\usepackage[symbol]{footmisc}

\begin{document}

	\title{Simultaneous single-qubit driving of semiconductor spin qubits at the fault-tolerant threshold}
	
	\author{W. I. L. Lawrie}
	\affiliation{QuTech and Kavli Institute of Nanoscience, Delft University of Technology, Delft, The Netherlands}
	\author{M. Rimbach-Russ}
	\affiliation{QuTech and Kavli Institute of Nanoscience, Delft University of Technology, Delft, The Netherlands}
	\author{F. van Riggelen}
	\affiliation{QuTech and Kavli Institute of Nanoscience, Delft University of Technology, Delft, The Netherlands}
	\author{N. W. Hendrickx}
	\affiliation{QuTech and Kavli Institute of Nanoscience, Delft University of Technology, Delft, The Netherlands}
	\author{S. L. de Snoo}
	\affiliation{QuTech and Kavli Institute of Nanoscience, Delft University of Technology, Delft, The Netherlands}
	\author{A. Sammak}
	\affiliation{QuTech and Netherlands Organisation for Applied Scientific Research (TNO), Delft, The Netherlands}
	\author{G. Scappucci}
	\affiliation{QuTech and Kavli Institute of Nanoscience, Delft University of Technology, Delft, The Netherlands}
	\author{J. Helsen}
	\affiliation{QuSoft and CWI , Amsterdam, The Netherlands}
	\author{M. Veldhorst}
    \email{Corresponding Author: M.Veldhorst@tudelft.nl}
	\affiliation{QuTech and Kavli Institute of Nanoscience, Delft University of Technology, Delft, The Netherlands}

	\maketitle

		\textbf{
Practical Quantum computing hinges on the ability to control large numbers of qubits with high fidelity. Quantum dots define a promising platform due to their compatibility with semiconductor manufacturing. Moreover, high-fidelity operations above 99.9\% have been realized with individual qubits, though their performance has been limited to 98.67\% when driving two qubits simultaneously. Here we present single-qubit randomized benchmarking in a two-dimensional array of spin qubits, finding native gate fidelities as high as 99.992(1)\%. Furthermore, we benchmark single qubit gate performance while simultaneously driving two and four qubits, utilizing a novel benchmarking technique called $N$-copy randomized benchmarking, designed for simple experimental implementation and accurate simultaneous gate fidelity estimation. We find two- and four-copy randomized benchmarking fidelities of 99.905(8)\% and 99.34(4)\% respectively, and that next-nearest neighbor pairs are highly robust to cross-talk errors. These characterizations of single-qubit gate quality are crucial for scaling up quantum information technology.   
		}

\section*{Introduction}
The reliable execution of quantum algorithms requires quantum processors that operate above the fault-tolerant threshold. This threshold depends on many aspects, but a 1\% error rate as evaluated in the surface code defines a common benchmark \cite{Raussendorf2007,Fowler2012}. The surface code requires a two-dimensional qubit system with nearest neighbour coupling, together with high-fidelity initialization and readout, and single and two-qubit gates. In both silicon and germanium \cite{Vandersypen2017a, Scappucci2020a}, two-dimensional approaches to scale-up quantum dots have been the focus of multiple recent efforts \cite{VanRiggelen2021, Ansaloni2020}. For electrons in silicon, scaling qubits in the second dimension is challenging due to the need for components such as striplines and nanomagnets to enable spin manipulation, limiting qubit realizations to linear arrays \cite{Veldhorst2015b, Takeda2020, Sigillito2019a, Philips2022}. Instead, spin qubits based on holes in germanium can be driven all-electrically through the intrinsic strong spin-orbit coupling \cite{Watzinger2018a, Hendrickx2020a, Hendrickx2020b}, allowing for very fast spin manipulation speeds that both increase the number of possible operations in a coherence time, as well as reduce the impact of low frequency noise on the qubit spin state \cite{Wang2022,piotSingleHoleSpin2022,Froning2021}. Furthermore, advances in strained germanium (Ge/SiGe) have yielded low charge noise and percolation density \cite{Lodari2021} and high hole mobility \cite{Sammak2019}, indicative of a highly uniform platform. These advantages have advanced the Ge/SiGe platform rapidly over the last few years and led to demonstrations of long spin relaxation times \cite{Lawrie2020a}, single hole qubits and singlet triplet qubits \cite{Hendrickx2020b, Jirovec2020} and universal operation on a 2x2 qubit array \cite{Hendrickx2021}.

However, as spin qubits expand into two dimensions, the growing number of possible qubit cross-talk interactions motivates careful characterization \cite{McKay2019}.
In the present work, we make use of a 2x2 quantum dot array of hole spin qubits, to characterize the single-qubit fidelities of the system, when driving one, two and four qubits simultaneously. We perform {single qubit randomized benchmarking} and investigate the dependence of fidelity on qubit {Rabi period (2$t_\pi$)}, finding that elementary gate single-qubit fidelities can be as high as 99.992(1)\%. We then investigate the individual single-qubit performance while benchmarking two and four qubits simultaneously, utilizing a novel randomizing benchmarking protocol called $N$-copy randomized benchmarking. We find two- and four-copy elementary gate fidelities as high as $F_{2Q}^{\pi/2}$ $=$  99.905(8)\% and $F_{4Q}^{\pi/2}$ $=$ 99.34(4)\% respectively. We compare these experiments at two magnetic fields $B_\text{ext}$ = 1 T and $B_\text{ext}$ = 0.65 T and find that while individual qubit operation is best at the lower magnetic field, simultaneous-qubit operation performs better at a higher magnetic field, due to the relevance of qubit addressability and qubit cross-talk. Finally we explore avenues for further improvement of qubit performance with respect to controlling the exchange interaction, preferred driving configurations, magnetic field amplitude and direction, and isotopic enrichment. 

\section*{Results}
\begin{figure*}
    \centering
    \includegraphics{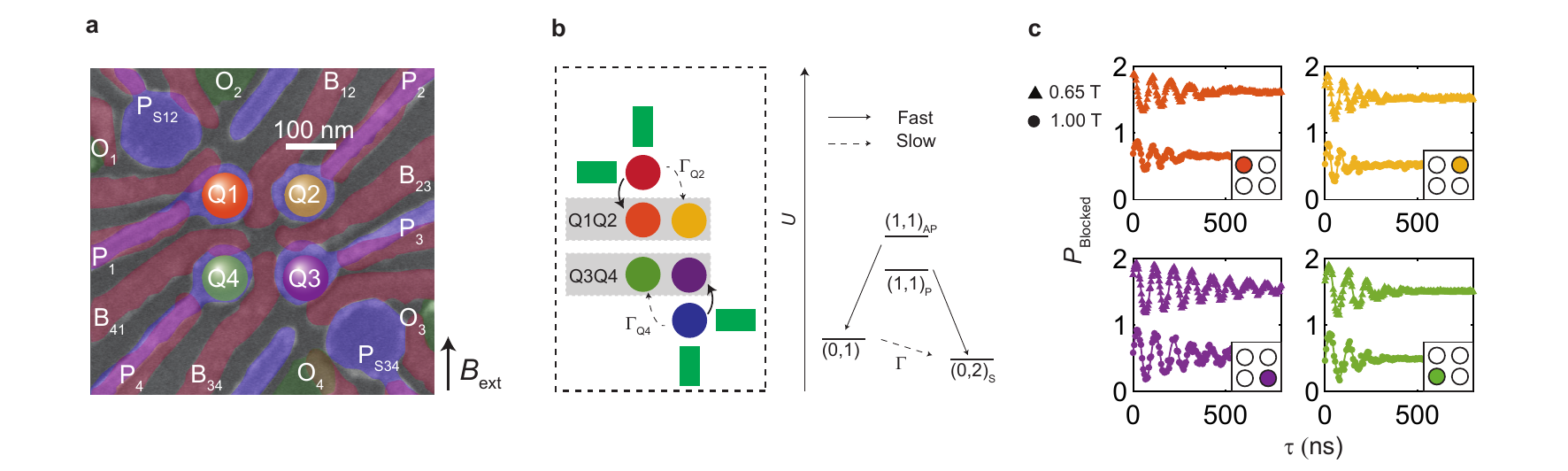}
    \caption{Heavy hole spin qubit array in germanium. (a) Scanning electron microscope (SEM) image of the device. The device is comprised of two lithographically defined layers of Ti:Pd constituting the plunger (blue) and barrier (red) gates, as well as an Al ohmic layer (green) forming low resistance ohmic contacts with the quantum well directly. An external magnetic field $B_{\text{ext}}$ is applied in-plane with respect to the quantum well. (b) Latched spin blockade readout mechanism. We consider two separate readout systems Q1Q2 (red) and Q3Q4 (blue), each containing a double quantum dot pair and single hole transistor. By reducing the reservoir tunnel coupling to quantum dot Q2(4) $\Gamma_{\text{Q2(4)}}$, we are able to suppress the (0,1) to (0,2)$S$ transition longer than the typical singlet triplet relaxation rates \cite{Hendrickx2020b} and facilitating a readout integration window of 10 $\mu$s. (c) Ramsey sequences on qubits Q1-4 respectively, at magnetic field $B_\text{ext}$ = 1 T (circles, lower) and 0.65 T (triangles, upper). Extracted spin dephasing times at 1 T for qubits Q1 (orange), Q2 (yellow), Q3 (purple) and Q4 (green) are $T_{2*,\text{Q1}}$ = 186 $\pm$ 19 ns, $T_{2*,\text{Q2}}$ = 119 $\pm$ 14 ns, $T_{2*,\text{Q3}}$ = 323 $\pm$ 52 ns and $T_{2*,\text{Q4}}$ = 147 $\pm$ 26 ns. At 0.65 T the dephasing times increase to $T_{2*,\text{Q1}}$ = 276 $\pm$ 22 ns, $T_{2*,\text{Q2}}$ = 166 $\pm$ 14 ns, $T_{2*,\text{Q3}}$ = 472 $\pm$ 31 ns and $T_{2*,\text{Q4}}$ = 228 $\pm$ 15 ns. Data for different magnetic fields is offset by unity for clarity. }
    \label{fig:main_1}
\end{figure*}

Figure \ref{fig:main_1}a shows a false-coloured scanning electron microscope (SEM) image of the device used in the experiment. The experimental setup details are identical to those found in Ref. \cite{Hendrickx2021}, excepting the qubit control configuration. It consists of two gate layers and an ohmic layer, where ohmic contacts to the quantum well are created by diffused Al \cite{Lawrie2020}. By applying potentials to the plunger gates P$_1$-P$_4$, we can define four quantum dots, each filled with a single hole spin such that we operate in the (1,1,1,1) charge regime (see Supplementary Note 1).  The tunnel couplings between these quantum dots can be tuned with the dedicated interdot barrier gates B$_{12}$-B$_{41}$. We also define two larger quantum dots using P${_{\text{S12}}}$ and P${_{\text{S34}}}$ in the multiple hole regime, which we utilize as charge sensors. By applying an rf tone to the ohmic gates O$_1$ and O$_3$ via two off-chip inductors bonded in-line, we form a resonant tank circuit allowing to perform fast rf charge sensing. To read out the spin states, we perform spin-to-charge conversion in the form of Pauli spin blockade (PSB). Figure \ref{fig:main_1}b depicts the two PSB readout pairs in the system, with the Q1Q2 system comprising of the two qubits Q1 and Q2 (orange, yellow), and the Q3Q4 system containing qubits Q3 and Q4 (purple, green). We make use of a latched readout mechanism \cite{HarveyCollard2018, Hendrickx2021}, whereby the dot-reservoir tunnel rate is limited significantly for one quantum dot per readout pair, which is depicted by a dashed arrow. We tune the dot-reservoir tunnel rates of quantum dots Q2 and Q4 to be $\Gamma_{\text{Q2}}$ = 5 kHz and $\Gamma_{\text{Q4}}$ = 0.416 kHz respectively, providing lower bounds on our readout bandwidth. We integrate in a readout window of 10 $\mu$s for both sensors such that we are well within this limit. {Microwave signals are applied as square pulses to the plunger gates, facilitating qubit state manipulation via electron dipole spin resonance of the qubits. In the present configuration, qubit Q1 is driven by plunger gate P4, qubits Q2 and Q4 are driven by plunger gate P2, and qubit Q3 is driven by plunger gate P3 (see Supplementary Notes 1 and 2 for details on the tuning).}

We apply an in-plane magnetic field $B_{\text{ext}}$ to split the spin states.{Two different magnetic fields were used in this experiment, of $B_{\text{ext}}$ = 1 T and $B_{\text{ext}}$ = 0.65 T, to provide a comparison between the different regimes of coherence, qubit response, and qubit resonance frequency spacing. (Fig. \ref{fig:main_1}c) shows how the spin dephasing time of qubit Q$_i$ ($T_{2*,Qi}$) performs at different magnetic fields, via Ramsey decay experiments. Here, $P_\text{blocked}$ is the readout probability of spin-blockaded readout. }

\begin{figure*}
    \centering
    \includegraphics{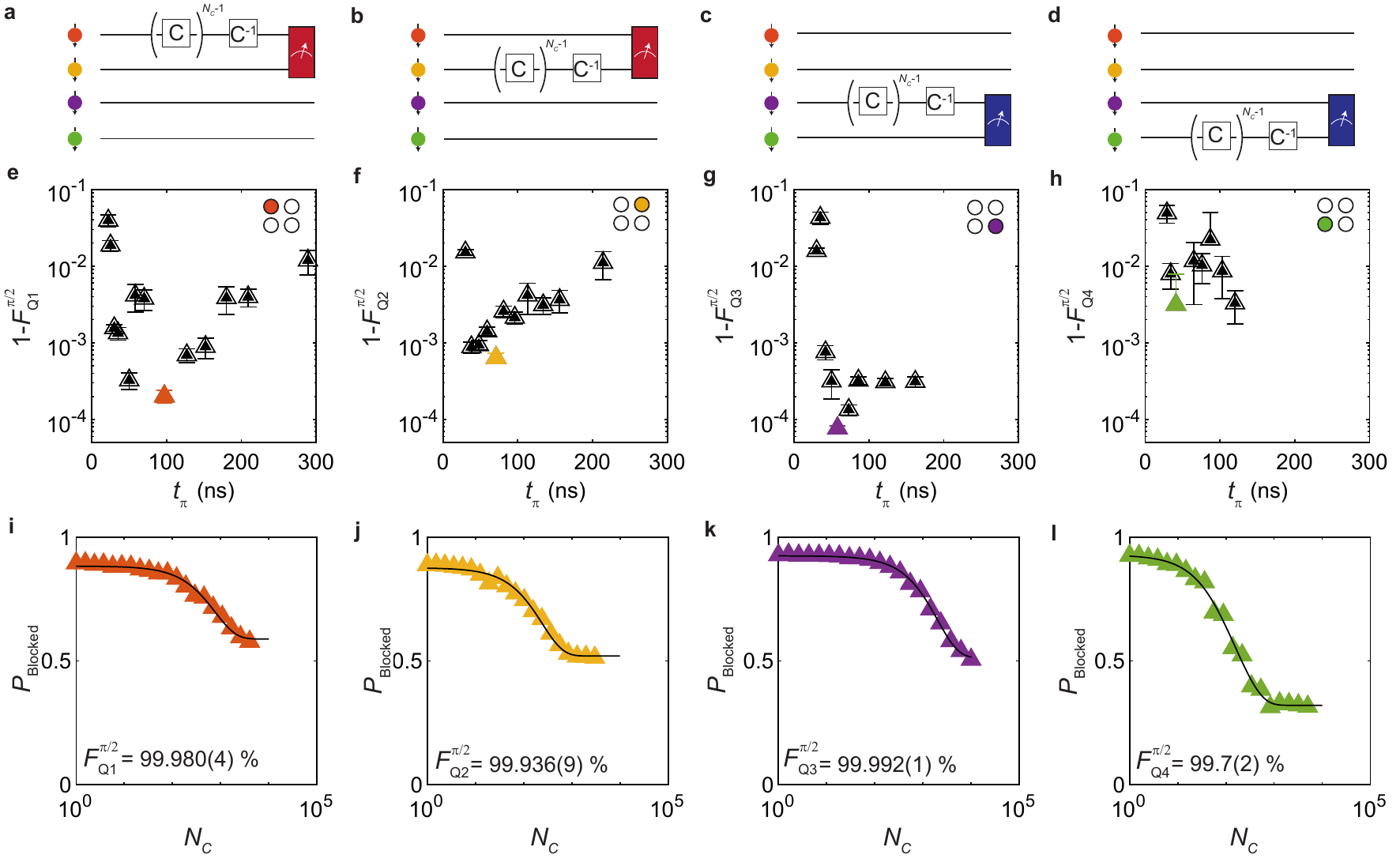}
    \caption{single-qubit randomized benchmarking at 0.65 T. (a-d) Random Clifford sequences applied to each qubit. Each qubit in the array is prepared in the spin down state. $N_C$-1 randomly selected Cliffords are applied to a single-qubit, after which a recovery Clifford (C$^{-1}$) is applied bringing the system back to the $|\downarrow\downarrow>$ state. Each sequence is repeated 32 times with different random permutations of Cliffords. Readout occurs via PSB on one of the two readout pairs Q1Q2 (red) or Q3Q4 (blue). (e-h) Dependence of qubit fidelity on $t_\pi$. An optimal $t_\pi$ occurs due to a trade-off between decoherence (high $t_{\pi}$) and errors introduced at low $t_\pi$ including gate calibration errors and driving non-linearities. Error bars reflect the fit uncertainty.(i-l) Best single-qubit benchmarks for each qubit. All native $\pi$/2 fidelities except $F^{\pi/2}_{\text{Q4}}$ exceed 99.9 \%, with $F_{\text{Q3}}^{\pi/2}$ exceeding four nines. These traces correspond to the respective highlighted points in (e-h).  }
   \label{fig:main_2}
\end{figure*}
\subsection*{Single-Qubit Randomized Benchmarking}
In order to get a baseline reading of the individual qubit fidelities, we performed randomized benchmarking to estimate the quality of each qubit. Randomized benchmarking provides the average fidelity of a gate set applied to  each qubit. Operations randomly selected from the Clifford group are applied to each single-qubit initialised in a known state. A final recovery Clifford C$^{-1}$ is applied to bring the qubit back to its original state. Imperfections in the applied gates and gradual qubit decoherence result in a decay of the recovered state probability as the number of applied Clifford operators is increased, allowing the extraction of a fidelity by fitting the decay \cite{Knill2008}. Each element of the Clifford group can be constructed from a variety of generator gates. We construct a Clifford group from a minimal generator set G${}_i$ $\in$ \{$ \text{X}_{\pi/2},\,\text{Y}_{\pi/2}$\} (see Supplementary Note 4). We find this set advantageous since it contains on average 3.217 qubit $\pi/2$ rotations (generators) per Clifford, that differ only by a software phase shift. This means the estimated Clifford fidelity is a direct indicator of the generator fidelity, by equally weighting the generators of the same length \cite{Cerfontaine2020}. Working with generators of equal length has the additional advantage of simplifying the experimental implementation of benchmarking qubits simultaneously \cite{Xue2019}.

Figures \ref{fig:main_2}a-d show the randomized benchmarking sequences for Q1-4 respectively. A red (blue) measurement window indicates PSB readout on the Q1Q2 (Q3Q4) double quantum dot pair. Each qubit is initialised in the spin down state by adiabatically pulsing the detuning through the (0,2)-(1,1) charge anticrossing. This is possible due to the large spin-orbit coupling for holes in Ge/SiGe, resulting in a coupling between the singlet (0,2)$S$ and the triplet (1,1)$T_-$ state \cite{Jirovec2020,Stepanenko2012,Hendrickx2021}. For each sequence length $N_C$, 32 random permutations of $N_C$ Cliffords are averaged to give the final trace, each of which comprises 1500 single shot measurements. An exponential decay is fit to the resulting trace (see methods), yielding a circuit level fidelity $F_{Q_i}$, from which an average generator fidelity $F_{Q_i}^{\pi/2}=1-(1-F_{Q_i})/(2\times3.217)$ can be extracted for each qubit. 
Holes in germanium allow for very fast electrical driving, with Rabi frequencies exceeding hundreds of MHz \cite{Wang2022,Froning2021}. However, rapid qubit manipulation is not always optimal for coherent qubit control, with high powers leading to enhanced systematic errors in qubit operation arising from effects such as sample heating or pulse imperfections. Indeed we find a strong dependence of the single-qubit fidelities on the drive speed. Figures \ref{fig:main_2}e-h show the generator infidelities (1-$F^{\pi/2}_{Q_i}$) as a function of qubit drive speed. Despite being able to drive qubit rotations in as fast as 10 ns, we find that the associated single-qubit fidelity suffers as a result, visible by a sharp decrease in the fidelity for qubits Q1 and Q3. Fidelity in these cases could be limited by a number of mechanisms, such as quantum dot anharmonicities 
\cite{Crippa2018,Scarlino2015} or systematic Pauli errors due to gate tuning.
{We also observe change in the resonance frequency of each qubit as a result of large applied microwave powers, where the single-qubit fidelity is observed to be lower.  }

{From the analysis we obtain that there is an optimum in the qubit driving speed (see Supplementary Note 3).} 

Figures \ref{fig:main_2}i-l show the randomized benchmarking data for the optimal $t_\pi$.  We extract generator fidelities above 99 \% for each qubit in the array, with qubit Q3 performing the best with $F{}_{\pi/2}^{\text{Q3}}$ = 99.992(1) \%, where the error on the last digit is given by the 95\% confidence interval of the fit uncertainty.

For single-qubit randomized benchmarking, we expect a fully decohered state to exhibit a blocked state probability of about $P_\text{Blocked}$ $\approx$ 1/2. However in the presence of finite exchange and classical cross-talk between the active qubit and the readout qubit in the spin blockade pair, state leakage can occur to all four states in the two-qubit subspace, resulting in a readout signal of about $P_{\text{Blocked}}$ = 0.329 (see Supplementary Note 1). We find that for the case of qubit Q4, the plateau of the spin blocked probability approaches the expected value of the fully depolarized two-qubit subspace for all driving powers.

This is likely due to the high power required to drive Q4 via plunger gate P$_2$ as a consequence of the larger distance between qubit and drive-gate, resulting in a large degree of cross-talk on qubit Q3. 
To account for state leakage, a second exponential decay is added for fitting randomized benchmarking traces for qubit Q4, yielding two characteristic decay constants (see methods). From this analysis, we calculate a {generator} fidelity $F_{\text{Q4}}^{\pi/2}$ = 99.7(2)\%, containing a leakage rate of $L^{\pi/2}_{\text{Q4}}$ = 0.07(2)\% per generator.
\subsection*{$N$-copy Randomized Benchmarking}
\begin{figure*}
    \centering
    \includegraphics{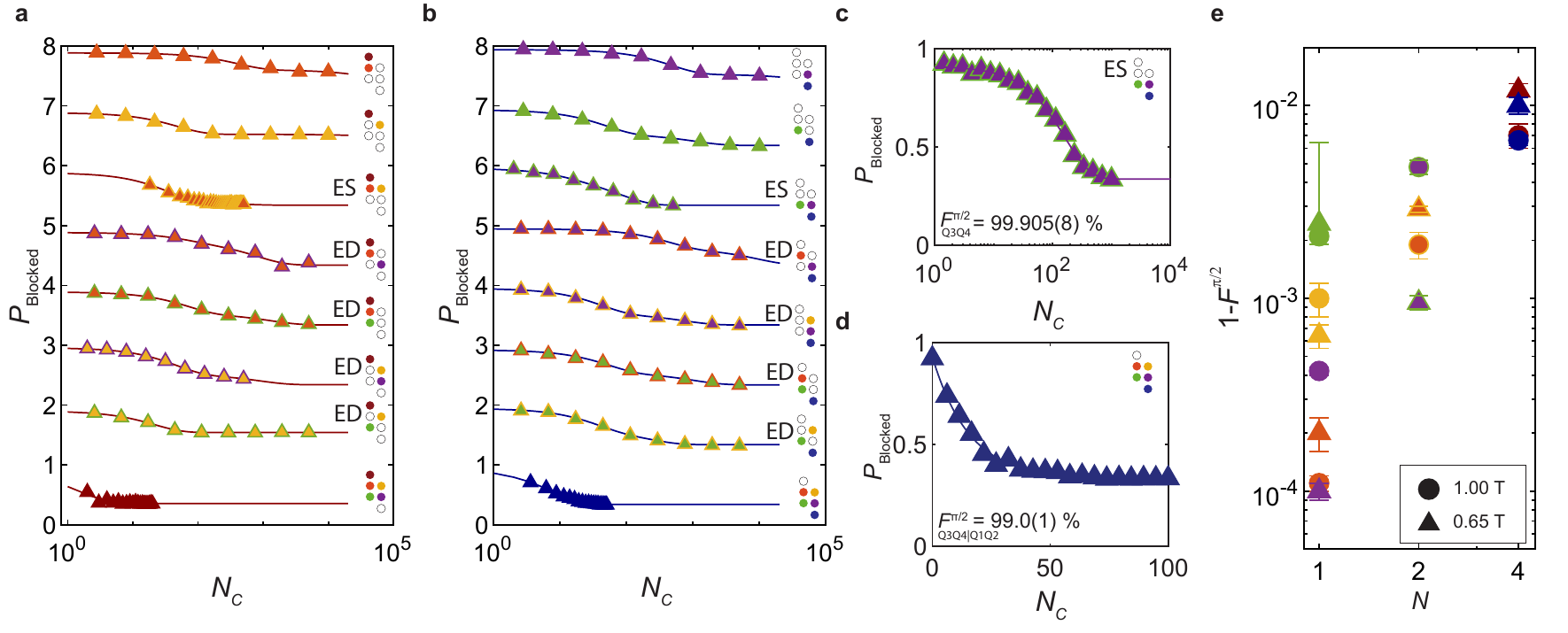} 
    \caption{{$N$-copy benchmarking in a four qubit array.} (a-b) A Two-copy benchmarking characterization at $t_{\pi}$ = 96 ns for all single-, two- and four-qubit configurations. (a) and (b) correspond to readout systems Q1Q2 and Q3Q4 respectively. For a pair of qubits, $N_C$-1 randomly selected Clifford sequences are applied simultaneously followed by a recovery pulse, and read out via PSB. Line and marker colour, are accompanied by cartoons indicating which qubits are being benchmarked, and which readout pair is being probed. The additional coloured dot indicates which system is being read out, such that a red(blue) dot indicates readout of the Q1Q2(Q3Q4) system. Fidelities and fit uncertainties of each trace are shown in Supplementary Note 6. (c) Decay for a two-qubit Two-copy benchmarking trace at 0.650 T. The readout system is Q3Q4. $t_\pi$ = 61 ns for the qubits. A simultaneous fidelity of $F_{Q3Q4}^{\pi/2}$ = 99.905(8) \% is extracted. (d) Four-qubits benchmarked simultaneously at 0.650 T. The readout system is Q3Q4. For all qubits, $t_\pi$ = 150 ns. A simultaneous fidelity of $F_{Q3Q4|Q1Q2}^{\pi/2}$ = 99.0(1) \% is extracted.  (e) {Summary of the best $N$-copy benchmarking results, reported as the native average $\pi/2$ generator infidelity 1-$F^{\pi/2}$ as a function of the number of qubits, $N$.} Data points for $B_\text{ext}$ = 0.65 T (triangles) and $B_\text{ext}$ = 1 T (circles) are compared. Colours indicate qubits involved in the experiment. {For $N$ = 1, red, yellow, purple and green represent the generator fidelities for qubits Q1, Q2, Q3 and Q4 respectively, in single qubit randomized benchmarking experiments. For N = 2 qubits, two-copy ES RB results are plotted. Orange/yellow markers symbolize qubits Q1 and Q2 generator fidelity for two-copy ES RB, while purple/green markers symbolize qubits Q3 and Q4 generator fidelity for two-copy ES RB. Red and blue markers for $N$ = 4 qubits represent all four qubits driven simultaneously, read out in system Q1Q2 and Q3Q4 respectively.} These fidelities are reported in tabular form alongside their Clifford fidelities, Rabi periods and magnetic fields in Supplementary Note 9. Quoted errors and error bars represent the fit uncertainty of the individual traces.}
    \label{fig:main_3}
\end{figure*}

Quantum algorithms will require the execution of a sequence of operations on a number of qubits, which need to be coherent for the duration of the algorithm. Sequential operation will significantly increase the run time of the algorithm and, crucially, may decohere idling qubits. Therefore, the operation of practical quantum computers will require multiple qubits to be controlled simultaneously\cite{Li2018a}. Doing so however can affect performance of the qubits due to qubit cross-talk. In conventional simultaneous randomized benchmarking \cite{Gambetta2012}, simultaneously driven qubit fidelities comprise the average of all available error channels for two or more qubits in the single qubit tensor product space. However, we envision that specifically pin-pointing cross-talk due to qubit manipulation will be of use for larger quantum systems with significant frequency crowding, as might be expected in spin qubit based quantum computing architectures \cite{Li2018a}. $N$-copy benchmarking is a variant of randomized benchmarking used to characterise the effects of qubit cross-talk in quantum processors with multiple qubits (see Supplementary Notes 5 to 9 for details). In contrast to simultaneous randomized benchmarking (SRB) \cite{Gambetta2012}, identical Clifford operations are applied to each qubit in the $N$ qubit sequence. This is advantageous for ease of experimental implementation, whereby each sequence is guaranteed equal duration, removing qubit idling time or Clifford sequence length compensation \cite{Xue2019}. 
 Depending on input state and measurement operator, different decay dynamics can be observed for two-copy randomized benchmarking experiments that can provide information about specific representations. By performing two-copy benchmarking on qubits in the same PSB readout pair, and an even-parity initial state (ES), for example, on qubits Q1 and Q2 starting in the state $\ket{\downarrow\downarrow}$, the decay structure simplifies to the sum of three real exponential decays (see Supplementary Note 7). An odd-parity input state on the same qubit readout system (OS) on the other hand, leads to a different decay profile, {comprising two real decays, and two complex conjugate decays} (see Supplementary Note 7) giving rise to a damped oscillatory dynamic (experimental implementation is shown in Supplementary Note 5). We note that performing both ES and OS measurements provides sufficient information about the different decay rates to rigorously lower bound the {simultaneous gate fidelity, that is, the average fidelity of single qubit gates applied to two qubits simultaneously, averaged over all single qubit Clifford gates}  (see Supplementary Note 8). The final decay structure of interest in this work, is the behaviour when performing two copy benchmarking on two qubits with an even parity initial state, in different readout systems (ED). Here, the decay structure is simplified to a single exponential decay (see Supplementary Note 7). We note that due to the different input states and measurement operators present in these two-copy randomized benchmarking experiments, as compared to a simultaneous randomized benchmarking experiment, the {decay rates} are not necessarily identical. {However, we show in Supplementary Note 7 that the two-copy ES RB protocol effectively yields the same decay curve as SRB for low exchange interaction strength, and is therefore a good estimator of the simultaneous gate fidelity for this experiment. For four-copy RB we do not have formal guarantees on the output, but it is well-behaved in practice and we report the resulting decay rates as four-qubit fidelities.  }

We now turn to $N$-copy benchmarking characterization of a 2x2 array of hole spin qubits. 
Here, the same generation procedure described in the single-qubit case is used to create a random sequence of single-qubit Cliffords, which is applied to all qubits in the experiment. The recovery Clifford always brings the system back to the $\ket{\downarrow\downarrow\downarrow\downarrow}$ state. 

Figures \ref{fig:main_3}a-d show a characterization of the single, two- and four-copy benchmarking at a consistent $t_{\pi}$ = 96 ns and magnetic field $B_\text{ext}$ = 0.65 T. From this dataset, we can approximate the relative loss in fidelity for each qubit due to the additional driving of another qubit (two-copy ED RB) by defining the quantity $\delta_{Q_i | Q_j}$ = $\epsilon_{Q_i|Q_j}$ - $\epsilon_{Q_i}$ where $\epsilon_{Q_i}$ = 1 - $F_{Q_i}$ is the infidelity of qubit Q$_i$ {as measured in a single qubit randomized benchmarking experiment,} and $\epsilon_{Q_i|Q_j}$ = 1 - $F_{Q_i|Q_j}$ is the infidelity of qubit Q$_i$ while simultaneously driving qubit Q$_j$ (Numerical data available in Supplementary Note 6). {This is measured by performing the two-copy ED RB experiment. We show in supplementary figure 6 that the extracted decay rate approaches simultaneous single-qubit gate fidelity ($\alpha_{i|j}$ in ref. \cite{Gambetta2012}) in the regime of zero qubit cross-talk.}

We observe that qubits driven alongside their nearest neighbours typically result in relatively larger error rates, while next nearest neighbour pairs can give error rates that are very low. For example, the relative loss of qubit Q3 due to the driving of qubit Q2 $\delta_{Q_3 | Q_2}$ = 0.35\% is significant, while for the non-adjacent qubit pair Q3 and Q1, $\delta_{Q_3 | Q_1}$ is immeasurable within the error bar and gave a simultaneous {single qubit generator} fidelity of $F^{\pi/2}_{Q3|Q1}$ = 99.964(3)\% $\approx$  $F^{\pi/2}_{\text{Q3}}$. The same is true for the opposite configuration, where $\delta_{Q_1 | Q_3}$ again is immeasurable within the error bar, and $F^{\pi/2}_{\text{Q1}}$ $\approx$ $F^{\pi/2}_{Q1|Q3}$ = 99.97(1)\%. The other next nearest neighbour pair Q2 and Q4 does not exhibit similarly low values of $\delta_{Q_2 | Q_4}$ or $\delta_{Q_4 | Q_2}$, however the driving geometry of this qubit pair is relevant (see Supplementary Note 6). 
Optimum $t_\pi$ values for simultaneous driving also exist due to a trade-off between qubit coherence and frequency crowding. The {four-copy generator decay rate} of the Q3Q4 system when all four qubits are driven simultaneously, $F^{\pi/2}_{Q3Q4|Q1Q2}$ = 99.0(1)\%, is optimal for $t_\pi$ = 150 ns (see Supplementary Note 6), while we find  $F^{\pi/2}_{Q1Q2|Q3Q4}$ = 98.76(2)\% for the Q1Q2 system at the same driving speed. Further increase of $t_\pi$ does not yield better fidelities as the coherence of qubit Q4 is limiting this value. We find we can also improve the simultaneous driving fidelity of the adjacent qubit pair $F^{\pi/2}_{Q3Q4}$ = 99.905(8)\% by decreasing $t_\pi$ to 61 ns, (see Fig. \ref{fig:main_3}c), indicating that decoherence may have be the limiting process.\\

The observation of high fidelity single-qubit gates for nearest and next-nearest neighbour qubit pairs, as well as fidelities above 99\% for four simultaneously driven qubits, shows promise for qubit operation in larger arrays.

Two different magnetic field settings were applied in this work to understand the importance of spin coherence and gate speed on the gate fidelities (see Fig. \ref{fig:main_1}c). At low magnetic fields, decoherence due to charge noise coupling in via the spin-orbit interaction is minimized, but also the Rabi speed is reduced. 

In order to retain the same $t_\pi$, the drive  power needs to be increased to compensate \cite{Terrazos2021}, leading to increased qubit cross-talk when driving simultaneously.

Figure \ref{fig:main_3}e summarises the randomized benchmarking results of this work, providing detailed information on the extracted average generator as a function of number of simultaneously driven qubits. On average, the individual single-qubit fidelities are higher at lower field, as are the results for driving two qubits simultaneously.
The power required to drive the systems at their optimal $t_\pi$ value however, was found to be inversely proportional with $B_\text{z}$ as expected, and as such the position of these points was also observed to shift to longer values of $t_\pi$. We observed that at the stronger magnetic field of 1 T, higher fidelities can be achieved for the case of four simultaneous driven qubits. We attribute this result to the faster achievable Rabi frequencies as well as larger qubit frequency splittings outweighing the relative loss in coherence. We note that the relative gain in four-copy qubit fidelity by operating at stronger magnetic fields is small, compared to the overall loss in fidelity with increasing number of simultaneously driven qubits. However, we expect that the magnitude of these classical cross-talk effects can be reduced through pulse engineering techniques \cite{Yang2019a,Barends2014}, facilitating shorter values of $t_\pi$ when driving qubits. Additionally, the high g-factor tuneability of hole states observed in Ge/SiGe \cite{Hendrickx2020b} could be exploited to reduce the cross-talk effect by maximising the separation of the qubits' resonance frequencies \cite{Arute2019a}.

\section*{Discussion}
While the individual and multi-qubit control results shown here define benchmarks for quantum dot qubit systems, we envision several strategies can be followed to further improve the fidelity. 
Precise control over the exchange interactions between adjacent qubits is extremely important for high fidelity quantum operations. In the current device we have a limited dynamic range over which we can control the exchange interaction. The overlapping gate structure and tight quantum dot definition \cite{Lawrie2020a} have proven essential in silicon and in particular SiMOS devices \cite{Veldhorst2015b}, but the low disorder and small effective mass of holes in germanium \cite{Lodari2019} enables extensive control over the exchange interaction \cite{Pan2020,VanRiggelen2021}. Therefore, relaxing the gate pitch may be feasible and the gate structure may thus be optimized to obtain larger on/off ratios for the exchange interaction. 

In the present work, we observe high single-qubit fidelities, using direct (Q2, Q3), nearest neighbour (Q1), and next-nearest neighbour (Q4) plunger driving strategies. For the first two, the relative power required to achieve similar driving speeds is comparable, and result in very high single-qubit fidelities, as well as no discernible state leakage. Driving Q$4$ via its next-nearest neighbour plunger P$_2$ however requires significantly larger powers to achieve comparable driving speeds, and results in larger qubit cross-talk to the direct and nearest neighbour qubits of P$_2$, resulting in clear state leakage. We also observe that the fidelity of Q$4$ is the lowest of the system. We therefore expect that an important parameter for optimal qubit control is thus the placement of the qubit driving gate. A particularly promising direction may be the engineering of barrier gates that can efficiently modulate the in-plane electric field. The device used here is based on a design where the barrier gates are patterned after the plunger gates and have a reduced pitch, which lead to a limited lever arm of the barrier gates. By widening the barrier gates and by patterning these gates before the plunger gates it may therefore be possible to both enhance the on/off ratio of the exchange coupling as well as to realise efficient qubit driving gates.
\\
The operation amplitude and angle of the external magnetic field is an important factor. In this work, we benchmark single-qubit gate fidelities at in-plane external magnetic fields of 0.65 T and 1 T. Spin dephasing times are higher at 0.65 T, however driving speeds and resonance frequency spacings are lower. We observe that trying to drive faster at lower field results in more frequency crowding due to the additional power required. This leads to larger cross-talk effects that ultimately limit fidelities when driving all qubits simultaneously, but allows for the highest single-qubit fidelities when driven sparsely owing to the enhanced coherence. Conversely, at higher field, we observe higher fidelities when driving simultaneously due to lower required powers, and higher frequency spacing, but lower fidelities when driving sparsely owing to the shorter coherence times.

Architectures targeting large-scale quantum computing with germanium can thus benefit from high-fidelity operation and fast and simultaneous quantum control. Pulse shaping techniques to overcome qubit cross-talk may be relevant in a dense qubit arrays, while the locality of the electric field may already be sufficient for sparse qubit arrays. Furthermore, germanium quantum technology offers a rich and diverse platform that can take advantage of integrated superconductivity for long-range links and where standard semiconductor technology may be incorporated \cite{Pillarisetty2011}, providing an exciting pathway for scalable and high-fidelity quantum technology \cite{Scappucci2020a}.

\section*{Methods}
\subsection{Device Fabrication}
The Ge/SiGe wafer is fabricated on a silicon substrate. We use reduced-pressure chemical vapour deposition to grow a 1.6 $\mu$m  strain-relaxed Ge layer, a 1 $\mu$m reverse graded Si$_{1-x}$Ge$_x$ (x varies from 1 to 0.8), a 500 nm constant composition Si$_{0.2}$Ge$_{0.8}$, and a 16 nm compressively strained Ge quantum well. Finally on the quantum well we grow a 55 nm Si$_{0.2}$Ge$_{0.8}$ barrier followed by an oxidised Si cap layer ($<$2 nm). An ohmic contact layer is created by first defining it using electron beam lithography, etching away the oxidised cap, then depositing 30 nm of Al. This layer is then covered in 7 nm of Al$_2$O$_3$ via atomic layer deposition at 300 $^o$C. The gate stack is in two overlapping layers of Ti:Pd (3/37 nm), separated by 7 nm of Al$_2$O$_3$.

\subsection{Generation of the Single-Qubit Clifford Set}
We quantify the quality of the single-qubit gates on all qubits by performing randomized benchmarking using the Clifford group $\boldsymbol{C}_1=\lbrace C_n\in U_1 | C_n \boldsymbol{P} C_n^\dagger = \boldsymbol{P}\rbrace $ with the single-qubit Pauli group $\boldsymbol{P}=\lbrace \textrm{I, X, Y, Z}\rbrace$. All 24 elements of the Clifford group are generated from a minimal set, $C_n=\prod_{g_i\in \mathcal{G}} g_i$ with $\mathcal{G}=\lbrace \textrm{X}_{\pi/2},\textrm{Y}_{\pi/2} \rbrace$. The average number of elementary gates per Clifford is 3.217. All 24 Clifford gates are provided in the Supplentary Note 4. While not the most efficient choice for elementary gates, this particular choice of set leads to uniform Clifford gates, whose elementary gates vary only by a $\theta=\pi/2$ phase-shift in the MW signal and are identical otherwise, beneficial for comparison and performing {$N$-copy benchmarking protocols.} 
\subsection{Fitting}
 Standard fitting of randomized benchmarking decays assumes a single exponential decay of the form $P_{\text{Blocked}}$ = $A$ $\times$ $F^{N_c}$ + $c$, where $A$ represents the visibility of the system, $F$ is the circuit level fidelity, $N_c$ is the number of Clifford sequences, and $c$ should be equal to the average signal of the $\ket{\downarrow,\downarrow}$ and $\ket{\uparrow,\downarrow}$ subspace (where the first index corresponds to the qubit being benchmarked). However, for qubit Q4, and for the high power regime of qubits Q1-3, we observe that the blocked signal plateaus to values corresponding to the fully decohered two-qubit subspace, indicating state leakage. In this case, we fit with two exponential decays $P_{\text{Blocked}}$ = $A_1$ $\times$ $(1-\epsilon)^{N_c}$ + $A_2$ $\times$  ($1-L$)$^{N_c}$ + $c$, where $c$ is set to the average signal of the four two-qubit states in the readout pair. Here, $\epsilon$ and $L$ correspond to the leakage-free and leakage error rates respectively, from which a circuit level fidelity can be extracted $F=1-\epsilon-L$ \cite{Andrews2019}. The estimate for leakage error rate is always taken to be the longer decay rate. {$N$-copy benchmarking may result in complex forms (see Supplementary Note 7). In particular, two-copy ES RB is sensitive to noise correlations. We have therefore performed numerical validation when using two-copy ES RB to obtain the average single qubit gate fidelity. Specifically, through numerical simulations we established that the experimental data can be fit with an exponential decay to obtain a metric that is representative of the average fidelity (see Supplementary Note 8).}
 
 For $N$-copy benchmarking involving qubits from the same readout pair, eg. Q1Q2 or Q3Q4, we fit a single exponential, giving a result that is representative of the average fidelity of both qubits in the system.
 
 For figure 3a and figure 3b, we fit a single exponential when the plateau corresponds to a fully depolarized single-qubit subspace, and a double exponential when it decays to a fully depolarized two-qubit subspace. In the latter case, the fidelity is always {approximated} from the faster of the two exponents. 
 
 \subsection{Simultaneous Elementary Gate Tuning}
 In order to tune the system for the case of two-simultaneously driven qubits, we first determine the non-simultaneously driven resonance frequency and Rabi frequency by applying a X$_{\pi}$ pulse to each qubit separately and perform readout out via PSB. We then tune the values of $t_{\pi}$ within 1 ns of each other and then set the desired simultaneous $t_\pi$ as the average of the two single-qubit values of $t_\pi$. This constitutes a rough calibration of the qubits. To fine tune, we apply a X$_{2\pi}$ rotation on both qubit simultaneously, and sweep the frequency of both qubits. Fitting a two-dimensional Gaussian then gives the first iteration correction to the two qubits' resonance frequencies. We then perform the same experiment, but varying the amplitude of the MW power of both qubits. When driving four qubits simultaneously, the tuning procedure is extended to include both systems Q1Q2 and Q3Q4 in the tuning iteration.
 
\subsection{Pulse Generation}
Microwave pulses are generated using IQ pulse modulation of a carrier frequency tone. Three mw signal generators (two Rhode Schwarz SGS100A, and one Keysight PSG-8267D) are used in the experiment, each with separate IQ and pulse modulation input channels from a Keysight M3202A AWG. At 0.65 T, the mw-signal generators operate with carrier frequencies 1.5 GHz, 2.1 GHz and 2.16 GHz, which route to plunger gates P4, P2, and P3 respectively. At 1 T, carrier frequencies are 2.3 GHz, 3.46 GHz, and 3.52 GHz. IQ frequency modulation can achieve a 300 MHz bandwidth around these carrier frequencies. All mw pulses in this experiment are square wave pulses.
 
 \subsection{Data Availability}
 The raw and processed data in this study are available in a Zenodo repository under accession code https://doi.org/10.5281/zenodo.7350228.
 
 \subsection{Code Availability}
 All scripts involved in data processing and analysis are available at https://doi.org/10.5281/zenodo.7350228.

\providecommand{\refin}[1]{\\ \textbf{Referenced in:} #1}

 \section*{Acknowledgements}
 We acknowledge Stephan Philips for software development, and Luca Petit for helpful discussions. M.V. acknowledges support through a Vidi grant, two projectruimte grants, and an NWA grant, all associated with the Netherlands Organization of Scientific Research (NWO). 
 J.H. is supported by the QSC Zwaartekracht grant (NWO). M.R. acknowledges support from NWO under Veni grant (VI.Veni.212.223). Research was sponsored by the Army Research Office (ARO) and was accomplished under Grant No. W911NF-17-1-0274. The views and conclusions contained in this document are those of the authors and should not be interpreted as representing the official policies, either expressed or implied, of the Army Research Office (ARO), or the U.S. Government. The U.S. Government is authorized to reproduce and distribute reprints for Government purposes notwithstanding any copyright notation herein.
 \section*{Author Contributions Statement}
W.I.L.L. fabricated the device and performed the experiments, with contributions from F.v.R and N.W.H. J.H. developed the theory and M.R. developed the numerical simulations. A.S. and G.S. supplied the heterostructures. S.L.d.S. contributed to software development. W.I.L.L and M.V. wrote the manuscript with input from all other authors. M.V. supervised the project.
 \section*{Competing Interests Statement}
The authors declare no competing interests.
\end{document}